\documentclass[preprint]{elsarticle}
\usepackage{graphicx}
\usepackage{dcolumn}
\usepackage{bm}
\usepackage{mhchem}
\usepackage[dvips]{color}
\usepackage{url}
\begin{document}

\begin{frontmatter}
\title{Principal Component Analysis of Azimuthal Flow in Intermediate-Energy Heavy-Ion Reactions}

\author{Bao-An Li\footnote{Bao-An.Li$@$tamuc.edu}}
\author{Jake Richter\footnote{jrichter2@leomail.tamuc.edu}}
\address{Department of Physics and Astronomy, Texas A$\&$M
University-Commerce, Commerce, TX 75429-3011, USA}
\date{\today}

\setcounter{MaxMatrixCols}{10}

\begin{abstract}
Principal Component Analysis (PCA) via Singular Value Decomposition (SVD) of large datasets is an adaptive exploratory method to uncover natural patterns underlying the data. Several recent applications of the PCA-SVD to event-by-event single-particle azimuthal angle distribution matrices in ultra-relativistic heavy-ion collisions at RHIC-LHC energies indicate that the sine and cosine functions chosen {\it a priori} in the traditional Fourier analysis are naturally the most optimal basis for azimuthal flow studies according to the data itself.  We perform PCA-SVD analyses of mid-central Au+Au collisions at $E_{\rm beam}/A$=1.23 GeV simulated using an isospin-dependent Boltzmann-Uehling-Uhlenbeck (IBUU) transport model to address the following two questions: (1) if the principal components of the covariance matrix of nucleon azimuthal angle distributions in heavy-ion reactions around 1 GeV/nucleon are naturally sine and/or cosine functions and (2) what if any advantages the PCA-SVD may have over the traditional flow analysis using the Fourier expansion for studying the EOS of dense nuclear matter. We find that (1) in none of our analyses the principal components come out naturally as sine and/or cosine functions, (2) while both the eigenvectors and eigenvalues of the covariance matrix are  appreciably EOS dependent, the PCA-SVD has no apparent advantage over the traditional Fourier analysis for studying the EOS of dense nuclear matter using the azimuthal collective flow in heavy-ion collisions.
\end{abstract}

\begin{keyword}
Equation of State, Heavy-Ion Reactions,Transport Models, Principal Component Analysis, Collective Flow
\end{keyword}


\end{frontmatter}
\newpage
\section{Introduction and conclusions}\label{S1}
A central goal of heavy-ion reaction experiments over a broad beam energy range from the Fermi energy all the way to LHC energies is to investigate the equation of state (EOS) of dense matter formed in these reactions. In realizing this goal, 
comparisons of hydrodynamics and/or transport model predictions with the experimental data of various components and/or forms of nuclear collective flow have been found very fruitful \cite{pawel85,oll,art}. In particular,  the analyses of single-particle azimuthal angle $\phi$ distribution $\frac{dN}{d\phi}$ with respect to the reaction plane have played an important role. Usually, a Fourier decomposition of the $\frac{dN}{d\phi}$ is performed according to
  \begin{equation}
\frac{2\pi}{N}\frac{dN}{d\phi} = 1 + 2\sum_{n=1}^{\infty}v_n\cos{[n(\phi-\Psi_{\rm PP_n})]}
    \label{eq:PP}
  \end{equation}
 where  $\Psi_{{\rm PP}_n}$ is the experimentally estimated azimuthal angle of the $n^{th}$ harmonic participant plane. The latter is normally taken as zero in model simulations where the true reaction plane is known. 
 The $v_n=<cos(n\phi)>$ is the $n$-th harmonic coefficient.  In particular, $v_1$ is the strength of the so-called directed flow and $v_2$ is that  of the elliptical flow. Since one can always do a periodic extension of the $\frac{dN}{d\phi}$ measured in some kinematic regions, the Fourier decomposition of  $\frac{dN}{d\phi}$ has been the standard technique for analyzing the azimuthal collective flow. 

While the sine and cosine functions constitutes mathematically a good basis for analyzing essentially all signals/observables, the question whether they are also naturally the most optimal basis according to the $\frac{dN}{d\phi}$ data itself was recently studied in Refs. \cite{Liu19,Alt19}. Interestingly, singular value decompositions of the particle azimuthal angle distributions generated by using the VISH2+1 hydrodynamic \cite{Liu19,hydro1,hydro2} and AMPT transport model simulations \cite{Alt19,AMPT} of ultra-relativistic heavy-ion collisions at LHC energies indicate that the leading principal component loadings (PCA eigenvectors in terms of the original observables) are naturally very similar \cite{Liu19} or almost identical \cite{Alt19} to the first few traditional Fourier bases. Moreover, it was found that mode-coupling effects are reduced for the flow harmonics defined by the PCA, indicating one of its possible advantages \cite{Liu19}. The systematics of heavy-ion reaction experiments from low to ultra-relativistic energies indicate that the strengths of both directed and elliptical flows reach their maxima in mid-central Au+Au reactions at beam energies around 1-2 GeV/nucleon \cite{Ha2}. A comparative study of collective flow using both the traditional Fourier and the PCA-SVD in heavy-ion collisions at these intermediate energies are thus useful. 

In this work, after first sorting and normalizing free nucleons into azimuthal angle bins (columns) in each event, we perform the PCA-SVD analyses of non-centered (normalized raw data), column-centered, and standardized (column-centered data scaled by the standard deviation in each column) $\frac{dN}{d\phi}$ data matrices generated for Au+Au collisions at $E_{\rm beam}/A$=1.23 GeV using the IBUU transport model \cite{libauer2,LiBA04}. We found that (1) in none of the analyses the PCs are naturally sine and/or cosine functions, (2) both the PC loadings and the corresponding singular values depend appreciably on the EOS used, (3) the singular value of the non-centered (normalized raw data) matrix is overwhelmed by the first PC reflecting merely the mean value of the nucleon azimuthal angle distribution while the PCs from the column-centered analysis reflect simply the decompositions of the standard deviations with their singular values decrease slowly. In addition, for the standardized data, there is basically no ``principal component'' as essentially all eigenvectors of the covariance matrix contribute approximately equally to the total covariance, and the PC loadings in different azimuthal angle bins show almost no correlation within large statistical fluctuations. For the purpose of investigating the EOS of dense matter using azimuthal collective flow in heavy-ion collisions around $E_{\rm beam}/{\rm nucleon}=1$ GeV, we conclude that the PCA-SVD approach has no advantage over the standard Fourier analysis.
  
The ultimate goal of the heavy-ion reaction community is to extract information about the EOS of dense nuclear matter. To help realize this goal and for the purposes of this work, we shall analyze the azimuthal flow in two ways: (1) the classic approach which explicitly uses the Fourier analysis and (2) the approach explained in Refs \cite{Liu19,Alt19} which use the PCA-SVD. To evaluate the importance of each approach in terms of its ability to reveal information about the EOS, it is necessary to first know where and how big are the EOS effects in the traditional analyses. While one can find qualitative answers about the EOS effects on directed and elliptic flows in the literature, we need specific and quantitative results from the traditional analyses to be compared with the PCA-SVD studies of the same reactions. For these reasons, we include in the Appendix results of the traditional azimuthal flow analysis to set proper references for comparisons. To make fair comparisons, we perform both differential and integrated flow analyses in the traditional approach. Knowing that the BUU model may not contain enough dynamical fluctuations, we also emphasize the importance of comparing results from the two approaches for the same sets of events generated by the same BUU code.

The rest of the paper is organized as follows. In section \ref{buu-model}, within the IBUU transport model we first examine the impact parameter and  EOS dependence of free nucleon azimuthal angle $\phi$ distributions $\frac{dN}{d\phi}$. We then outline in Section \ref{approach}
the PCA via SVD formalism and data pre-processing procedure. Since in the PCA-SVD literature, different approaches have been widely used in pre-processing the raw data leading to PCs and the corresponding singular values having different meanings, we perform our PCA-SVD analyses using non-centered (normalized raw data), column-centered, and standardized $\frac{dN}{d\phi}$ matrices. Results of these PCA-SVD analyses will be presented in Section \ref{pca-ana}. Finally, we summarize.
  
 \section{IBUU transport model predictions for free nucleon azimuthal angle distributions in mid-central Au+Au collisions at $E_{\rm beam}/A$=1.23 GeV}\label{buu-model}
 In studying nuclear collective flow in heavy-ion reactions at intermediate-relativistic energies, Boltzmann-Uehling-Uhlenbeck (BUU)-like transport models \cite{Sto86,Bertsch,Cas90,Xu19,Maria} played a particularly important role for extracting useful information about the EOS of dense matter \cite{das93,res97,Bass,Pawel02}. We use here an isospin-dependent BUU model \cite{libauer2,LiBA04}. Most of its details and many applications are reviewed in Ref. \cite{BALI}. For the purposes of this work, we use the simplest momentum independent isoscalar single-nucleon potential corresponding to an incompressibility of K=230 MeV (soft) and K=380 MeV (stiff), respectively. The Coulomb potential and a simple symmetry potential ($F_3$ in Eq. (3) of Ref. \cite{Li97}) are also used. For the comparative studies here, this choice is sufficient and computationally efficient. While the more advanced potentials with momentum dependence for both the isoscalar and isovector single-nucleon potentials are more physical \cite{LiChen05}, the large number of reaction events necessary for the present study is unfortunately computationally prohibitive. We also only look at free nucleons identified as those nucleons with local densities less than $1/8$ the saturation density $\rho_0$ of nuclear matter in the final state of the reaction. We notice that the collective flow signatures in Au+Au collisions at $E_{\rm beam}/A$=1.23 GeV have been studied recently by the HADES Collaboration \cite{Hades,Ha2} using free protons and light clusters. To compare quantitatively with the HADES data would require us to use momentum-dependent single-nucleon potential and a coalescence model coupled to our transport model. Such a study is planned. 

In our analyses, nucleon azimuthal angle distributions $\frac{dN}{d\phi}$ are calculated with respect to the true reaction plane $(x-o-z)$ of the simulations with the beam along the $z$ direction. Making use of the symmetry of the reaction system considered, we calculate the $\frac{dN}{d\phi}$ for $0\leq \phi\leq \pi$ with the $\phi={\rm arccos}(p_x/p_t)$ where $p_t=(p_x^2+p_y^2)^{1/2}$ is the transverse momentum. Since the angle $\phi$ obtained this way is independent of the sign of $p_y$, the $\frac{dN}{d\phi}$ obtained is a result of averaging its distributions in $0\leq \phi\leq \pi$ and $\pi\leq \phi\leq 2\pi$. 
While reducing the statistical errors, this way of calculating the azimuthal angle may reduce or disallow some high-order correlations, such as the back-to-back emission and the triangular emission pattern. Nevertheless, for the stated purposese of this work, the chosen method of calculating the $\phi$ distribution in $0\leq \phi\leq \pi$ is sufficient and we have checked that the main conclusions of our work are qualitatively independent of this choice.
 
\begin{figure*}[thb]
\centering
\includegraphics[width=0.3\textwidth]{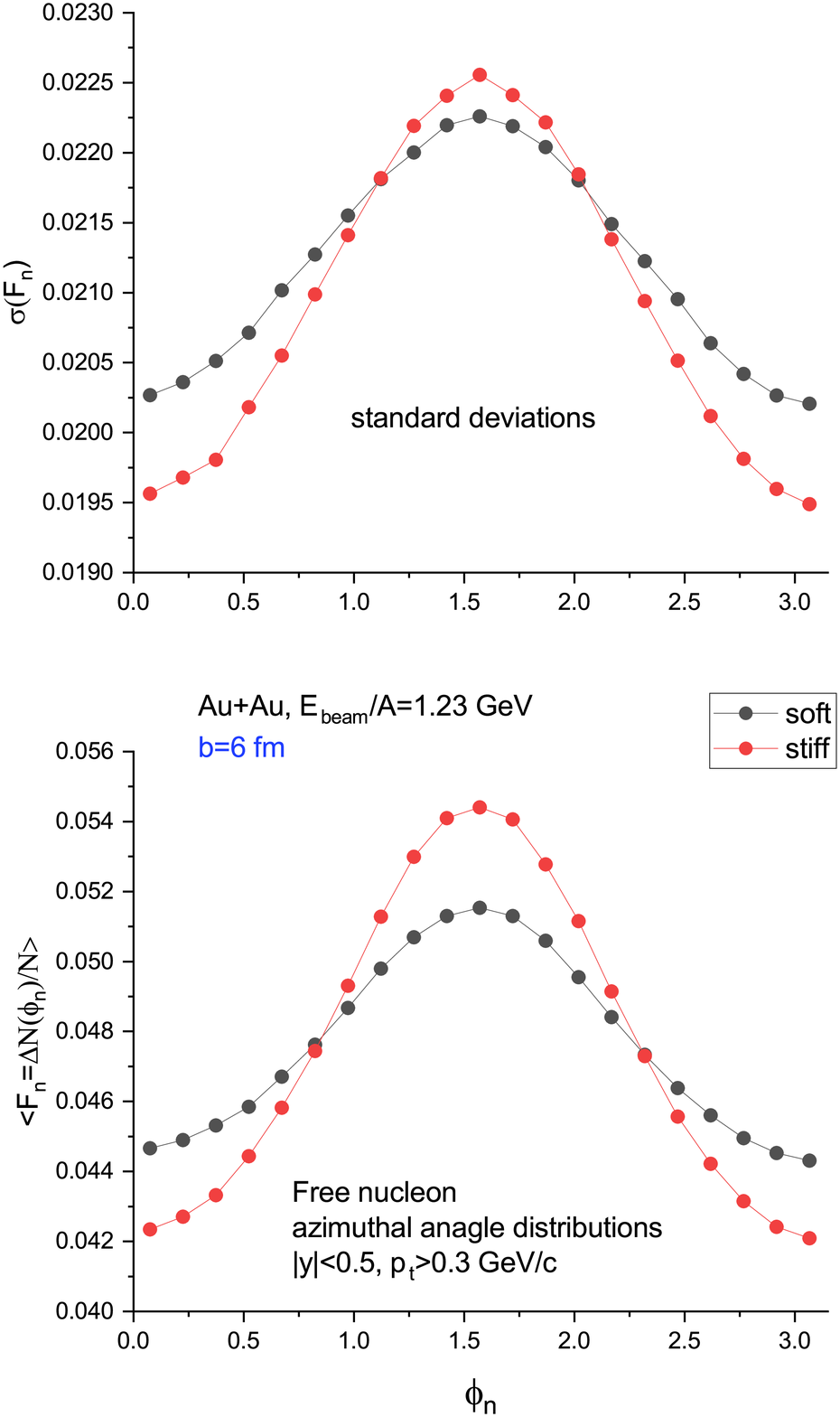}
\includegraphics[width=0.3\textwidth]{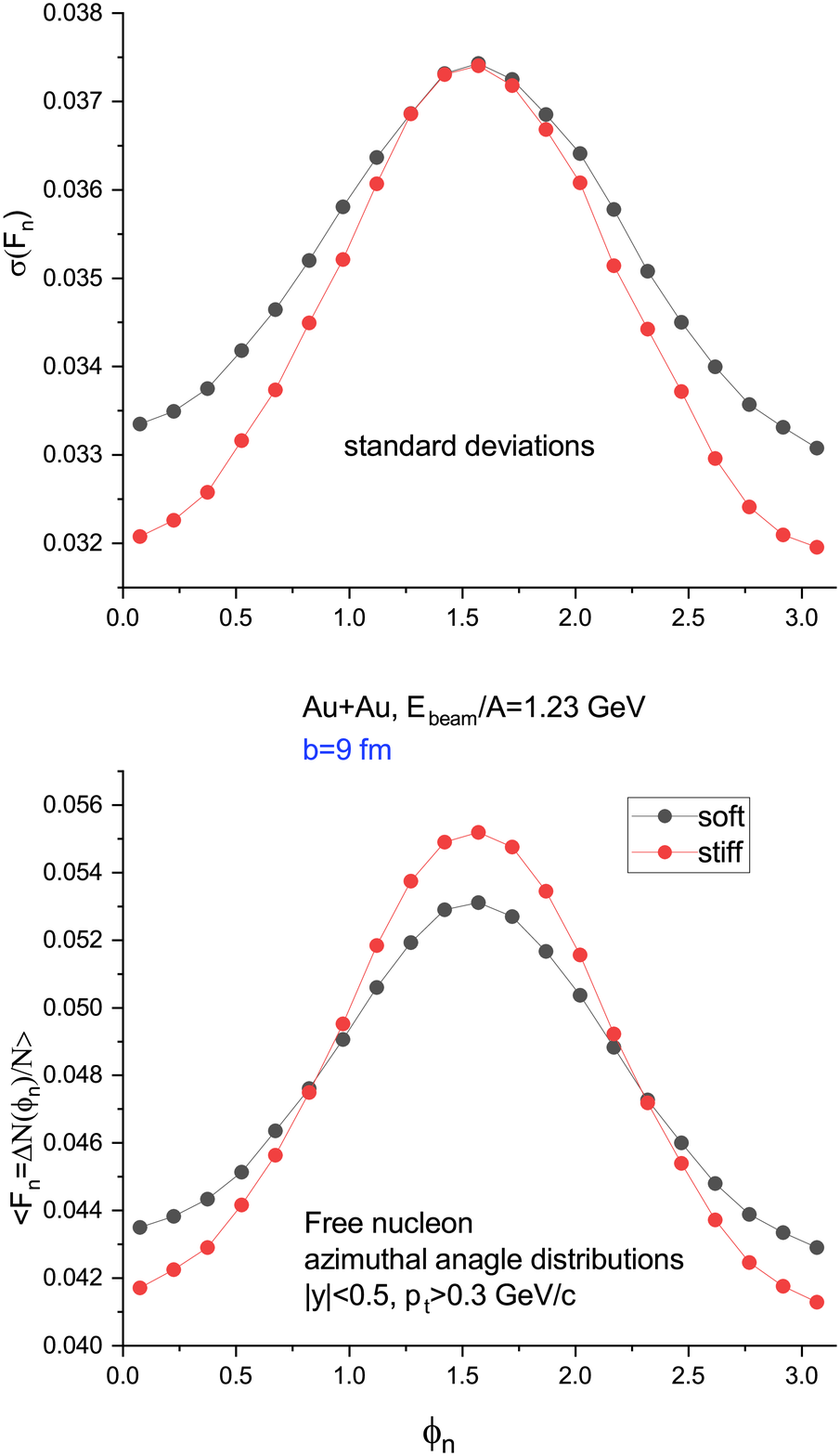}
\includegraphics[width=0.3\textwidth]{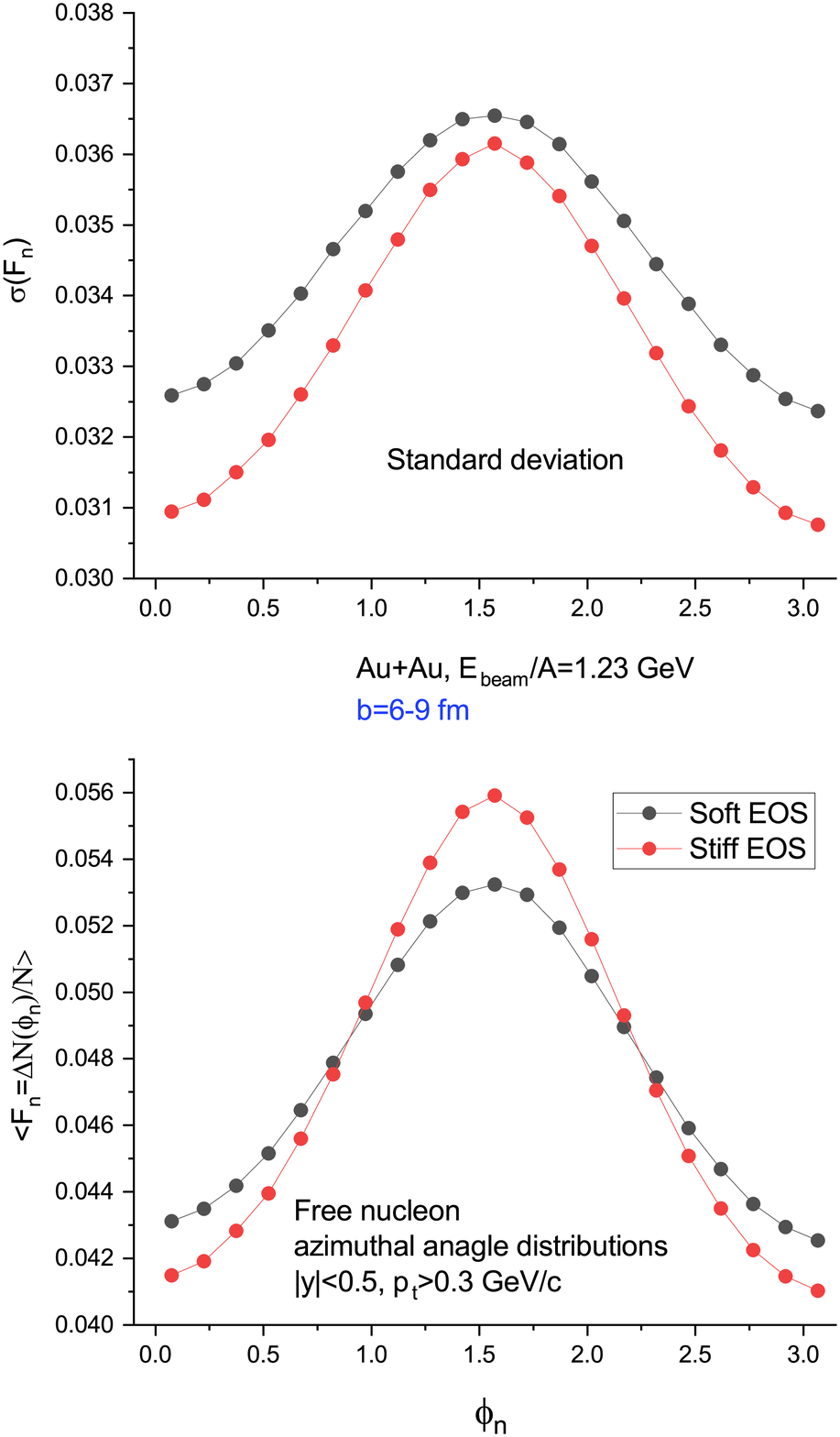}
\caption{Azimuthal angle distributions of the normalized fractions of free nucleons in Au+Au collisions at $E_{\rm beam}/A$=1.23 GeV with a soft (K=230 MeV) and a stiff (K=380 MeV) EOS, respectively. Lower panel: mean values. Upper panel (note the change in scale for the 3 cases): standard deviations from 0.5M events with an impact parameter of b=6 fm (left), 9 fm (middle) and 1.5M events with b between 6 fm and 9 fm (right), respectively, for each EOS.} \label{dndf}
\end{figure*}
To get rid of effects due to trival fluctuations, such as the total number of free nucleons that is fluctuating from event to event especially in mixed datasets from reactions with different impact parameters that can not be fixed in experiments, we first normalize the total number of particles (e.g., free nucleons detected in a particular rapidity- transverse momentum region) to be analyzed in each event to one as normally done in preparing dataset for PC analyses. Specifically, we define $F_n(\phi_n)\equiv \Delta N(\phi_n)/N_{\rm free}$ where $\Delta N(\phi_n)$ is the number of free nucleons in the n-th $\phi$ bin of size $\pi/21$ and $N_{\rm free}$ is the total number of free nucleons in the particular event in the kinematic region considered. Here we use 10 bins on each side of the central bin at $\pi/2$. We have checked that our conclusions do not change but having larger statistical fluctuations if we double the bin number.  

\begin{figure}[thb]
\centering
\includegraphics[width=0.55\textwidth]{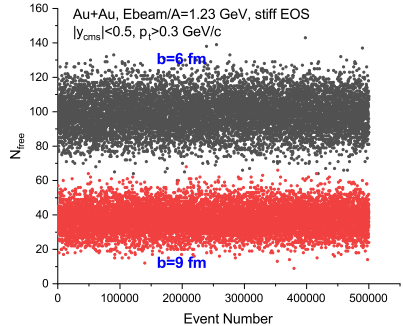}
\caption{Total numbers of free nucleons in the specified kinematic region of Au+Au reactions as functions of event number.}\label{nmax}
\end{figure}

The effectiveness of getting rid of the trivial fluctuations using the $F_n(\phi_n)$ can be seen in Fig. \ref{dndf}, where the mean values and standard deviations of $F_n(\phi_n)$ for free nucleons with $|y_{\rm cms}|\leq 0.5$ and $p_t\geq 0.3$ GeV/c in reactions with b=6 fm, 9 fm and mixed events with b between 6 and 9 fm, are shown,  respectively. As we shall discuss later, the separate evaluation of the means and standard deviations are useful for the PCA via SVD of the particle azimuthal angle distributions. The mean values and standard deviations of $F_n(\phi_n)$ all peak around $\phi=\pi/2$ indicating clearly an elliptical flow pattern. It is seen that the azimuth asymmetry(e.g., $<F_n(\pi/2)>/[<F_n(0)>+<F_n(\pi)>]$) measured by the $\phi_n$ dependence of $<F_n(\phi_n)>$ is appreciably stronger with the stiff EOS as one expects. As shown in the Appendix, with the stiff EOS the elliptical flow $v_2$ is larger and there are more free nucleons (larger stopping power) in the mid-rapidity region than the case with the soft EOS. Most interestingly, the mean $F_n(\phi_n)$
distribution is approximately independent of the impact parameter although the absolutely number of free nucleons are very different as shown in Fig. \ref{nmax}. While the standard deviations of $F_n(\phi_n)$ is significantly larger with a larger impact parameter (e.g., b=9 fm) when the number of free nucleons is smaller, as one expects from statistics. The mixed event is dominated by reactions with larger impact parameters, it thus has a standard deviation close to that with b=9 fm.

Experimentally, the impact parameter of a reaction can not be fixed. Data analyses using various centrality triggers select events within some impact parameter ranges. To see the size of particle number fluctuations due to the impact parameter selection, shown in Fig. \ref{nmax} are the total numbers of free nucleons in the specified kinematic region of Au+Au reactions as functions of event number with a ``fixed" impact of b=6 fm and b=9 fm, respectively. Technically, in all calculations, we used 500 test-particles/nucleon in each reaction with a fixed impact parameter. In generating the 0.5 million (M) b=6 fm or b=9 fm events we used 1000 random impact parameters between 6.0 (9.0) fm and 6.1 (9.1) fm, respectively.  Results of PCA-SVD analyses of these events with approximately fixed impact parameters will be compared with those from analyzing 1.5M mixed events with b generated randomly between 6 fm and 9 fm. It is seen that the $N_{\rm free}$ is strongly impact parameter dependent. Quantitatively, it scatters between about 20 to 130 in mixed reactions with b between 6 and 9 fm.

The results shown above clearly indicate that the BUU model is able to generate large event-by-event fluctuations even for reactions with ``fixed impact parameters". It has been a longstanding problem for the transport model community to know to what extents the sizes of these fluctuations are big enough and if their forms are physical. One of the challenges has been the lack of proper tools and relevant data for investigating the nature and calibrating the size of fluctuations. The number of particles $\Delta N(\phi)$ in bin $\phi_n$ or its normalized fraction $F_n(\phi_n)\equiv \Delta N(\phi)/N_{\rm free}$ is often regarded in the literature as a one-body observable while the correlations between the populations of two bins as a two-body observable. But, the one-body here actually contains many particles. Thus, even the one-body observable contains information about the multi-particle dynamics governed by the BUU equation involving a mean-field term and a collision integral. The latter is a major source of dynamical fluctuations from event to event. In the BUU model, one uses the test-particle approach \cite{Wong} in evaluating the mean-field. As a result, some events share the same mean-field but different collision integrals, i.e., they are not completely independent. This may lead to an under evaluation of dynamical fluctuations from event to event depending on the relative importance of the mean-field and collision integrals as well as some technical/numerical aspects of the codes \cite{Herman}. Thus, the possible under evaluation of dynamical fluctuations in the BUU approach in simulating heavy-ion reactions depends on the beam energy of the reaction. At low energies below and around the Fermi energy, the mean-field dominates the dynamics while at higher energies the collisions becomes gradually more important. At relativistic energies, the dynamics is dominated by the collision integrals. Thus, the general criticism of lacking dynamical fluctuations in BUU calculations should be applied conditionally. On the other hand, it is well known that both the mean-field and collisions are important for generating correlations/collectivity/flow in individual events or on the event averaged basis in heavy-ion collisions at beam energies from Fermi energy to several GeV/nucleon. The BUU approach has been very successful in extracting useful EOS information from flow data in these reactions analyzed in the traditional approach as demonstrated in the Appendix.
 
 \section{PCA via SVD formalism and data pre-processing}\label{approach}
The PCA has been widely used in many fields of sciences and engineering. It reduces the dimensionality of large datasets by creating new uncorrelated variables that successively maximize variance. 
Since the new variables are defined by the dataset itself instead of {\it a priori} PCA is an adaptive data analysis tool \cite{PCA-review}. There are many textbooks and articles about the fundamentals and applications of PCA in the literature. Here we adopt the terminologies from the recent review \cite{PCA-review} on PCA via SVD using column-centered, non-centered (raw data) and standardized data matrices. For discussions on the relationship and advantages/disadvantages of these three ways of preparing the data matrices we refer the readers to Ref. \cite{PCA-raw}. 
In applying the PCA via SVD to the azimuthal angle distributions of particles in heavy-ion collisions, we follow the approach used in Refs. \cite{Liu19,Alt19}. 

According to the PCA via SVD formalism \cite{PCA-review}, any arbitrary matrix $\mathbf{Y}$ of dimension $n\times p$ can be decomposed with three matrices according to 
\begin{eqnarray}
\quad\quad\quad\quad \mathbf{Y}=\mathbf{{U}{L}{A}^T}
\end{eqnarray}
where $\mathbf{{U}}$ and $\mathbf{{A}}$ are $n\times r$ and $p\times r$ ($r\leq {\rm min(n,p)}$) matrices with orthonormal columns while $\mathbf{{L}}$ is a $r\times r$ diagonal matrix with decreasing singular values ${\sigma}_j$ (j=1 to r). The covariance matrix $\mathbf{S}$ of the data is given by 
\begin{eqnarray}
\quad\quad\quad\quad (n-1)\mathbf{S}=\mathbf{{Y}^T{Y}}=\mathbf{{A}{L}^2{A}^T}.
\end{eqnarray}
The columns of $\mathbf{{A}}$ are the eigenvectors of $\mathbf{{Y}^T{Y}}$ (thus also $\mathbf{S}$), while those of $\mathbf{{U}}$ are the eigenvectors of $\mathbf{{Y}{Y}^T}$ 
and $\mathbf{L}^2$ is a diagonal matrix with the squared singular values (the eigenvalues of $(n-1)\mathbf{S}$). 

Adopting the approach and notations used in Refs. \cite{Liu19,Alt19} in applying the PCA-SVD to azimuthal angle distributions of particles from ultra-relativistic heavy-ion collisions, we can sort particles into $m$-bins (columns) in azimuthal angle $\phi$ for $N$ number of reaction events (rows). The elements $m_{i,j}$ of the resulting data matrix $\mathbf{M_f}$ of dimension $N\times m$ are the number of particles in the $i$-th event (raw) and $j$-th $\phi$ bin (column) with $i$ from 1 to $N$ and $j$ from 1 to $m$. Applying the SVD to $\mathbf{M_f}$, one can write \cite{Liu19} 
\begin{equation}\label{svd-0}
\quad\quad\quad\quad \mathbf{M}_f=\mathbf{{X}{\Sigma}{Z}}=\mathbf{{V}{Z}}.
\end{equation}
We notice that $\mathbf{Z}=\mathbf{{A}^T}$, thus the rows (columns) of the $\mathbf{Z}$ $(\mathbf{{A}})$ matrix contains the loadings (coefficients) of the principal components (PCs) in terms of the original variables. The azimuthal angle distribution in the $i$-th event $dN/d\phi^{(i)}$ can be expressed by the linear combination of the eigenvectors $z_j$ (the $j_{th}$ row of matrix $\mathbf{Z}$) with $j=1,2,... ,m $ as \cite{Liu19} 
\begin{equation}\label{svd-1}
dN/d\phi^{(i)}=\sum_{j=1}^m {x}_j^{(i)}{\sigma}_j {z}_j
=\sum_{j=1}^m \tilde{v}_j^{(i)} {z}_j.
\end{equation}
If the singular value $\sigma_j$ decreases quickly with $j$, normally the first few PCs will be sufficient to account for most of the covariance of the $\mathbf{S}$ matrix. In this case, the above summation can be truncated at $k$ significantly less than $m$. The $\tilde{v}_j^{(i)}$ can be further averaged over all events to evaluate on average how each PC contributes to the event averaged azimuthal angle distribution $dN/d\phi$ (We shall use the phrase ``Event averaged PC coefficients" in presenting the event averaged $\tilde{v}_j^{(i)}$ in the following). 
\begin{figure*}[thb]
\centering
\includegraphics[width=0.32\textwidth]{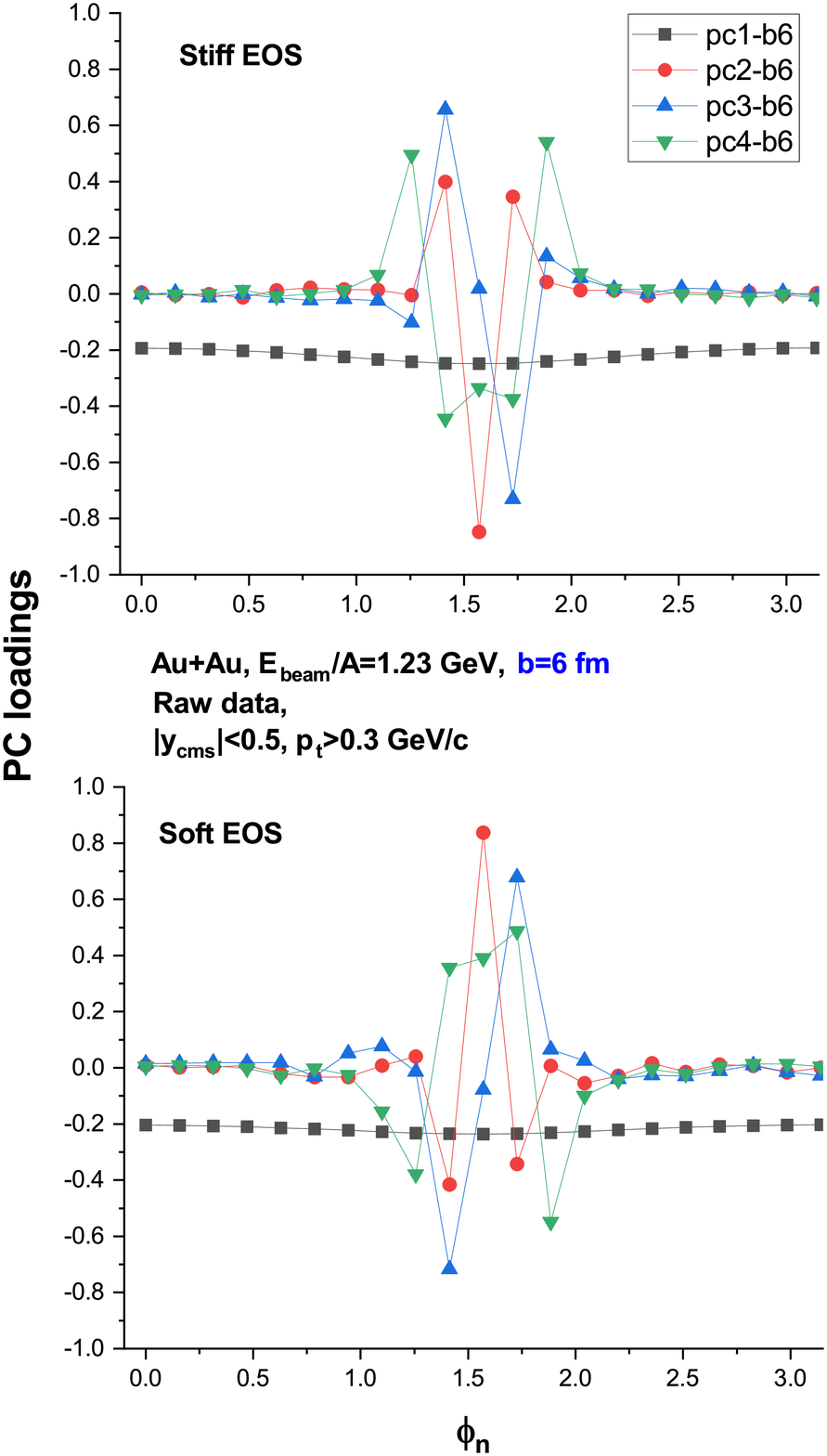}
\includegraphics[width=0.32\textwidth]{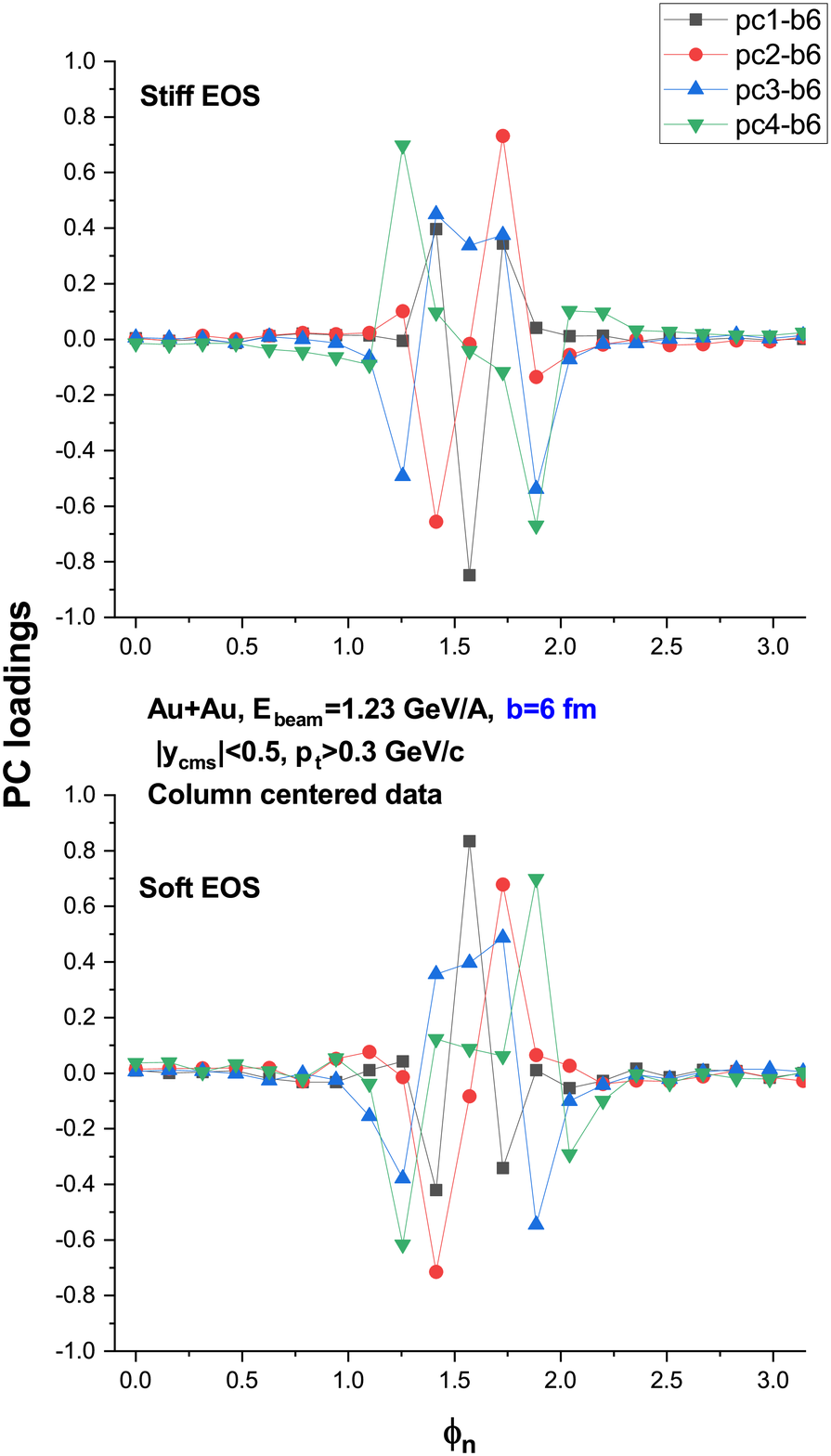}
\includegraphics[width=0.32\textwidth]{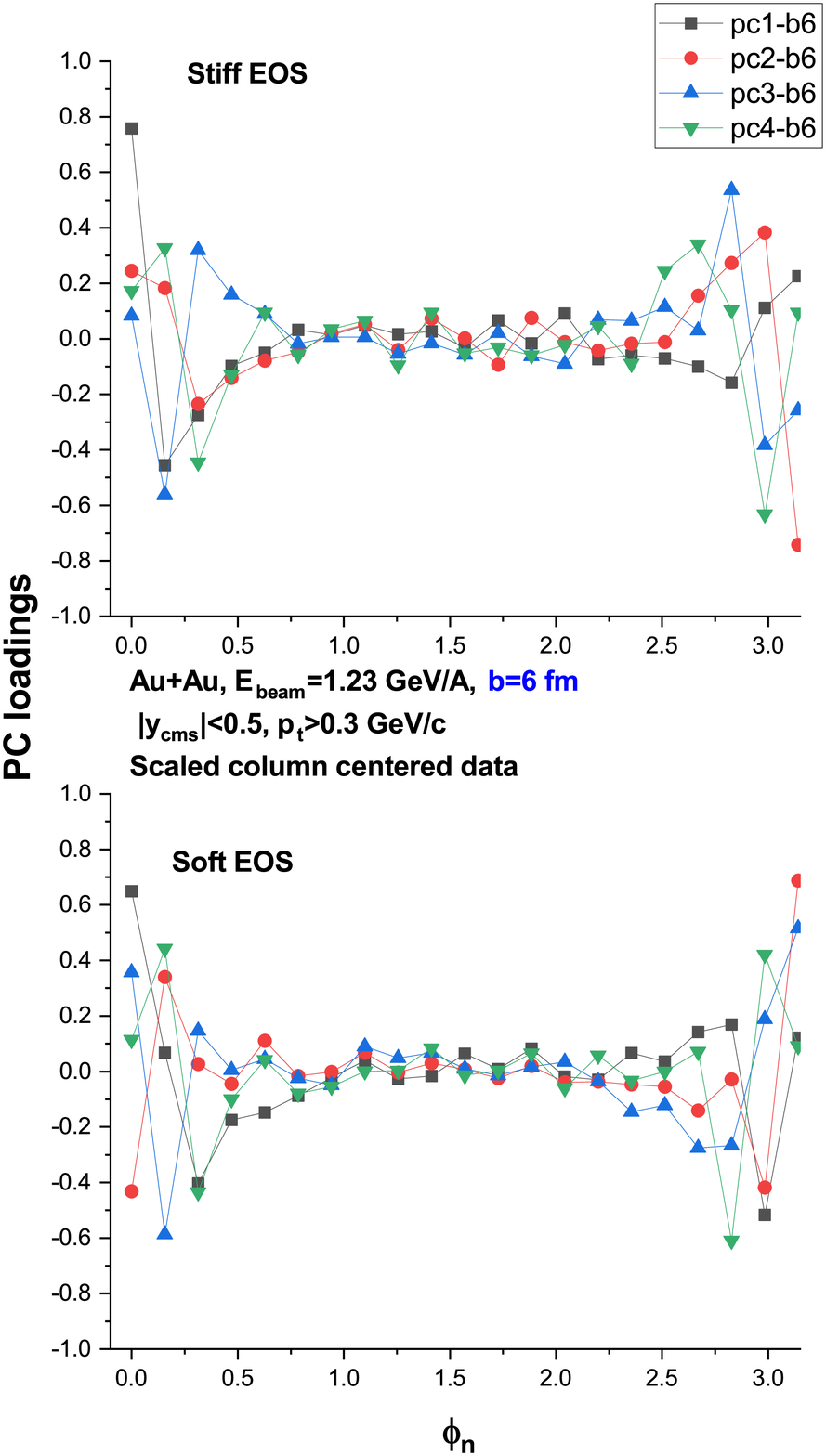}
\caption{PC loadings from the 0.5M events of Au+Au collisions with an impact parameter of 6 fm using the raw (left), column centered (middle) and standardized (right) data matrices, with the stiff (upper) and sfot (lower) EOS, respectively}\label{loading1}
\end{figure*}

To this end, it is necessary to compare the nature of the tranditional flow analyses we carrried out in the Appendix with the PCA-SVD outlined above. It is clear and well understood that all quantities, i.e., the average in-plane transverse momentum $< p_x/A >$, the  strength of directed flow $v_1=<cos(\phi)>$, and the strength of elliptical flow $v_2=<cos(2\phi)>$ used in the traditional flow analyses are all event averaged. It is well known that while BUU-like models may lack to some extent dynamical fluctuations from event to event depending on the in-medium particle-particle scattering cross sections, they are among the most successful models in describing the collective flow within the traditional approach \cite{Sto86,Bertsch,Cas90,Bonasera,Xu19,Maria,das93,res97,Bass,Pawel02,Li-sustich}. A key reason is that all the quantities studied are event averaged. It is also well known that both the mean-field and particle-particle collisions are important for reproducing the experimentally measured strengths of collective flow. 
What about the PCA-SVD analysis of the event-by-event matrices of the particle fractions in the azimuthal angle bins? While all the three major outputs, namely, the PC loadings (eigenvectors), singular values (eigenvalues) and the ``Event averaged PC coefficients"  $\tilde{v}_j^{(i)}$ are results of event averaging, they do carry information about the event-by-event fluctuations of the particle azimuthal angle distribution. 
Each matrix element of the covariance matrix $\mathbf{S}$ is evaluated from 
$S_{i,j}=E[(F_i-E[F_i])([(F_j-E[F_j])]$ where the operator E denotes taking the expectation value of its argument using the distributions of $F_i$ and $F_j$ that are fractions of particles in the azimuthal angle bin $i$ and $j$, respectively. The $S_{i,j}$ thus carry information about the event-by-event fluctuations of $F_i$ and $F_j$. 
The patterns of PC loadings of the covariance matrix represent how the fluctuations and/or means of $F_i$ are correlated from bin to bin depending on how the $F_i$ is prepared. They reflect the collectivity of nucleons at different levels of importance measured by their corresponding singular values.

In the literature, different ways have been used in preparing the data matrix $\mathbf{M_f}$ \cite{PCA-review}. For the azimuthal angle distribution, it is first normalized by the total number of particles $N$ detected in the kinematic region considered to obtain the normalized raw data $dN/d\phi/N$. We refer the corresponding data matrix $\mathbf{M}_f$ as the raw non-centered data matrix. Its elements are $m_{i,j}$ in index notation. If one subtract from $m_{i,j}$ its event averaged value $<m_j>$ in each column ($\phi$ bin), namely $m_{i,j}-<m_j>$, the resulting data matrix is the so-called column-centered matrix. Furthermore, if one divides the $m_{i,j}-<m_j>$ with its standard deviation $\delta_j$ in each $\phi$ bin, one obtains the standardized data matrix with elements $[m_{i,j}-<m_j>]/\delta_j$. Using the means and standard deviations of $dN/d\phi/N$ in each $\phi$ bin shown in Fig. \ref{dndf} we constructed the above three kinds of data matrices. 

The advantages and disadvantages of using the three matrices for the PCA-SVD analyses as well as the interpretations of the resulting PCs were discussed using several examples in Refs. \cite{PCA-review,PCA-raw}. We also notice that in the literature there are debates on whether the Gaussian distribution of the dataset is required or not \cite{Pca1,Pca2}. According to Ref. \cite{PCA-review}, PCA as a descriptive tool needs no distributional assumption. Indeed, PCA has been used on various data types. We notice that Ref. \cite{Liu19} used the normalized raw data matrix while the column-centered data matrix was used in Ref. \cite{Alt19} in the PCA analyses of RHIC/LHC reactions. While we have no physics reason to expect the standardized data matrix to show any harmonic behavior, we perform our PCA analyses using all three matrices to evaluate effects of the data pre-processing. 

It is also important to emphasize that PCA has been successfully used previously in studying event-by-event fluctuations, fine structures and couplings of various flow components/forms in ultra-relativistic heavy-ion collisions \cite{Bha15,Maz15,Maz16,Sir17,Boz18}. These studies analyze covariance matrices constructed by using two-particle correlations in rapidity-transverse momentum bins. As pointed out already in Refs. \cite{Liu19,Alt19}, these studies all assume that the single-particle azimuthal angle distribution is a harmonic function. They also use different normalization schemes. The covariance matrices using two-particle correlations contain more information about details of nuclear collective flow compared to the ones built on correlations between populations in two azimuthal angle bins as done here and in Refs. \cite{Liu19,Alt19}. 
\begin{figure*}[thb]
\centering
\includegraphics[width=1.0\textwidth]{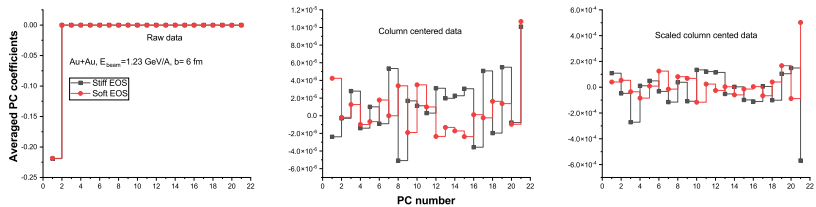}
\caption{The event averaged PC coefficients for the three data matrices analysed.}\label{b6CE}
\end{figure*}
\begin{figure}[thb]
\centering
\includegraphics[width=0.45\textwidth]{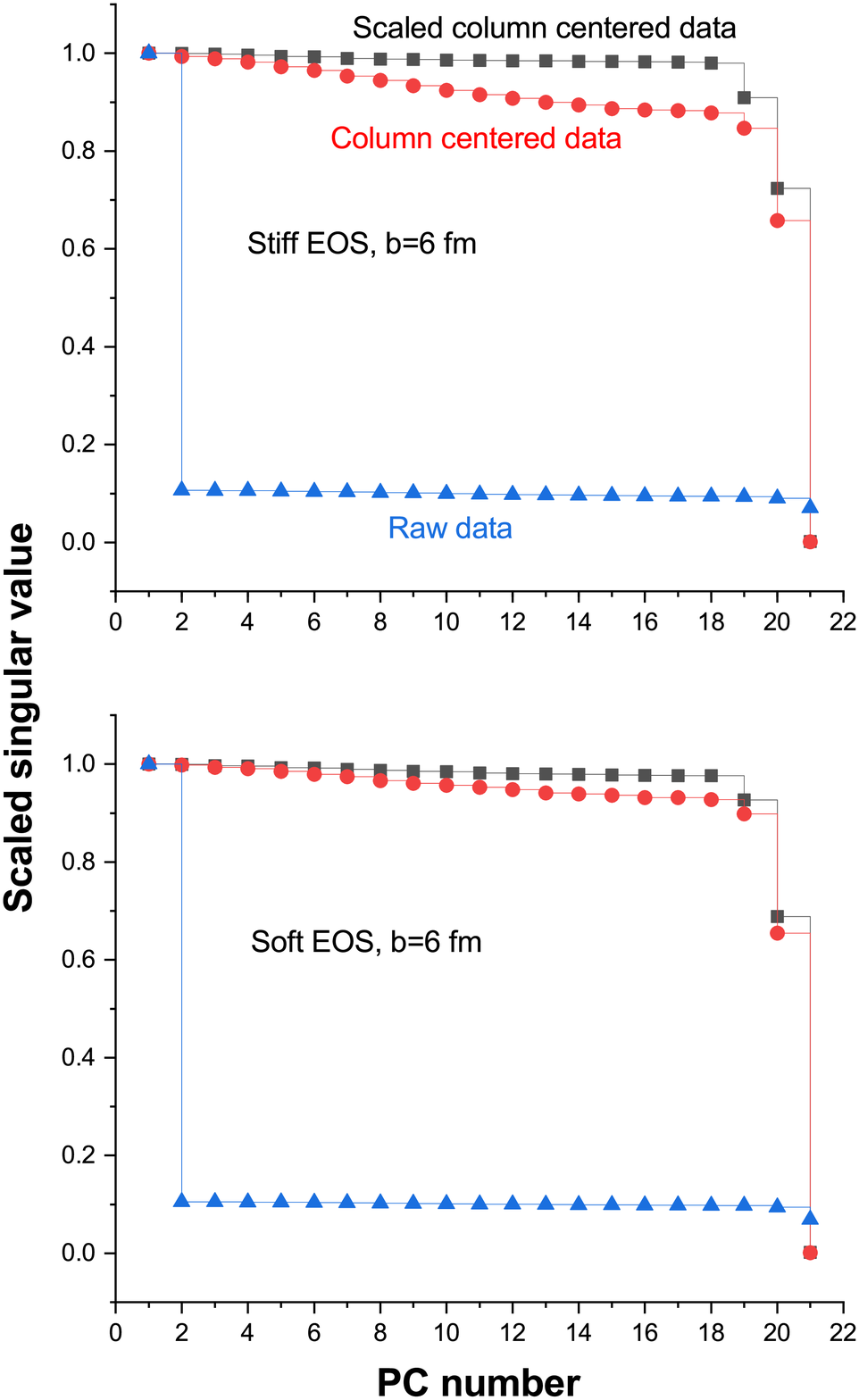}
\caption{The scaled (with respect to the singular value of PC1) singular values as functions of the number of principal components for the three data matrices considered using the stiff (upper) and soft (lower) EOS, respectively.}\label{b6SV}
\end{figure}

\section{Singular value decompositions of nucleon azimuthal angle distributions in mid-central Au+Au collisions at $E_{\rm beam}/A$=1.23 GeV}\label{pca-ana}
In the following, we compare the first few PC loadings, the singular values and the event averaged PC coefficients obtained by using the three ways of preparing the data matrices.  We notice that since the eigenvectors can be multiplied by a minus sign without changing any physical content, only the pattern and relative signs of the PC loadings in a given analysis are relevant. We use the default sign conventions used in the PCA package \cite{pca-site} built on sklearn which is a robust machine learning library in Python \cite{sklearn}.

\begin{figure*}[thb]
\centering
\includegraphics[width=0.32\textwidth]{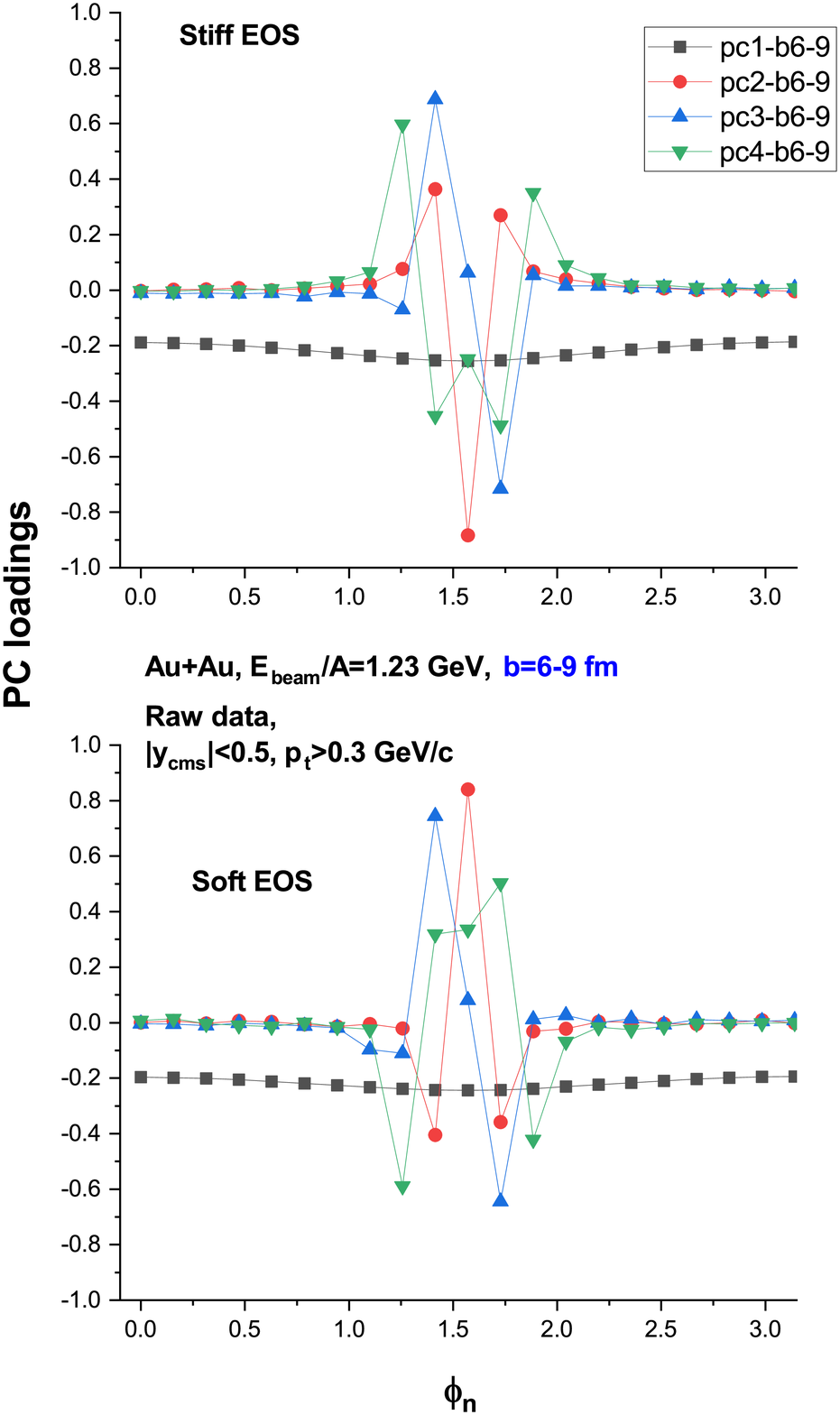}
\includegraphics[width=0.32\textwidth]{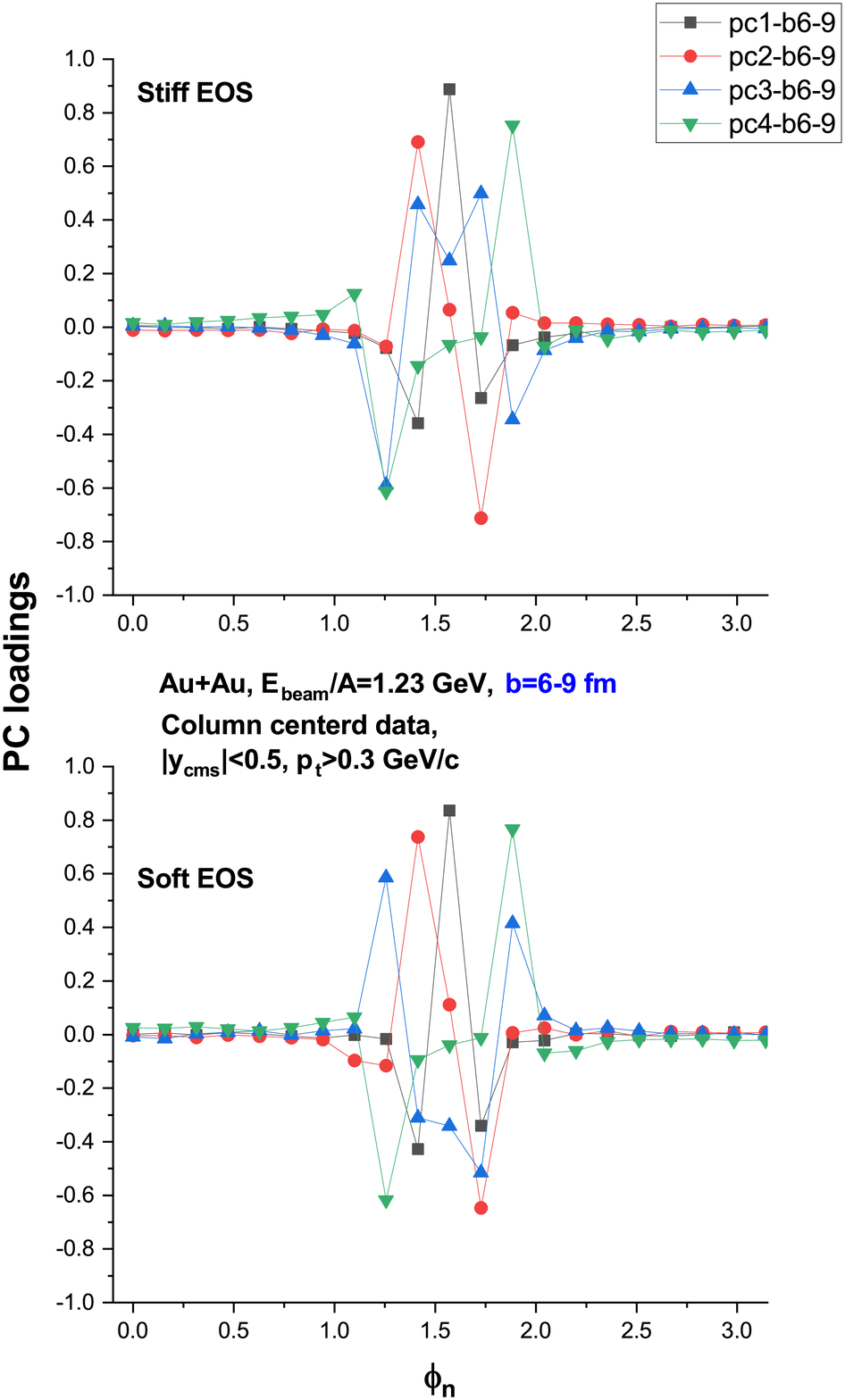}
\includegraphics[width=0.32\textwidth]{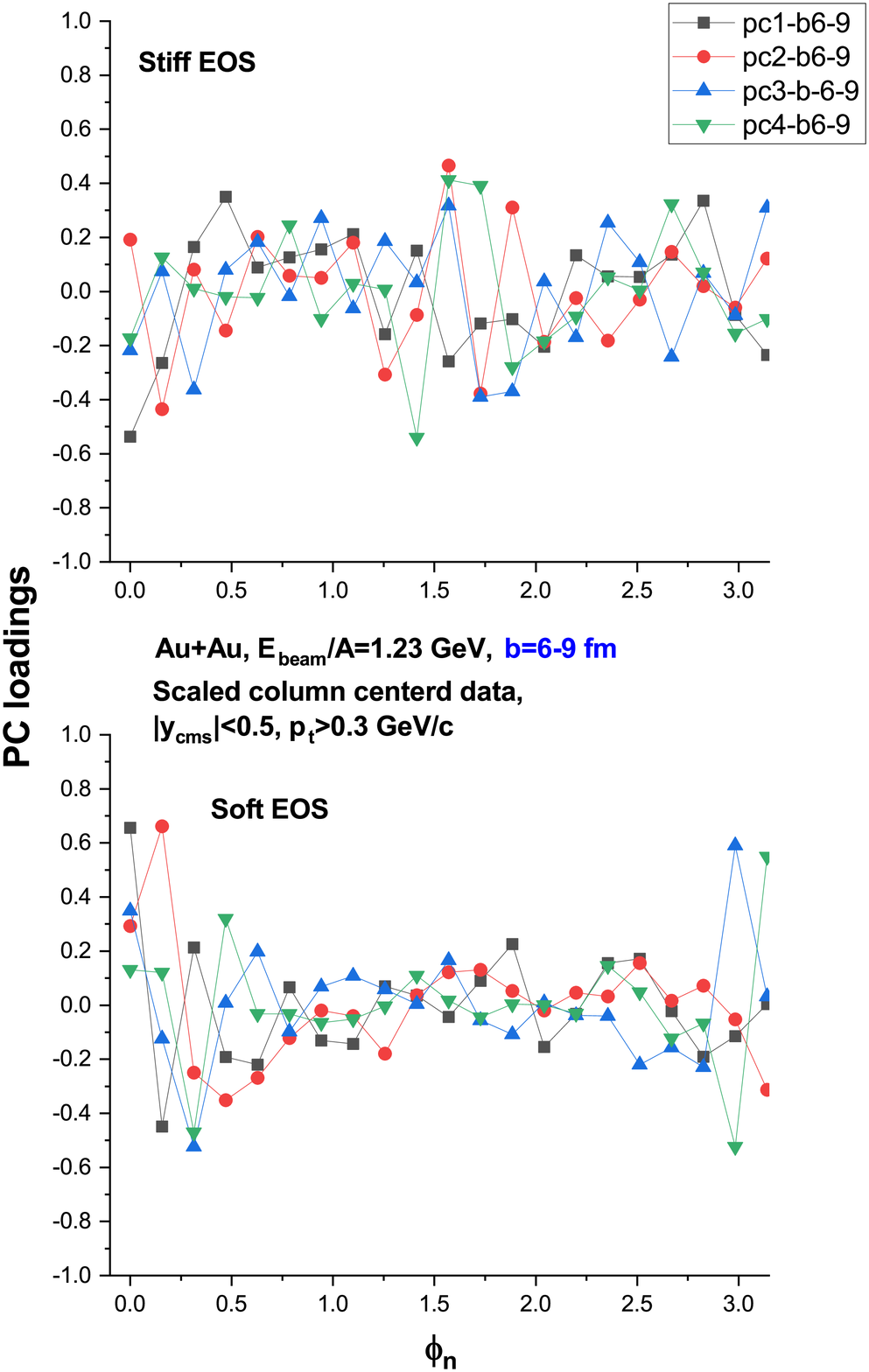}
\caption{Same as in Fig. \ref{loading1} but for 1.5M mixed events with impact parameters between 6 fm and 9 fm.}\label{bmixed}
\end{figure*}

Shown in Fig. \ref{loading1} are PC loadings of the first 4 principal components for the 0.5M events of Au+Au collisions at $E_{\rm beam}/A$=1.23 GeV with an impact parameter of b=6 fm and the stiff (upper) or soft EOS (lower) for the raw data (left panels), column centered (middle panels) and standarized (right panels) data, respectively. With the raw data (i.e., normalized but not column centered), the PC1 loading (black squares in the left panels) resembles the mean value of $dN/d\phi/N$ shown in Fig. \ref{dndf} except a minus sign. Its dependence on the EOS is appreciable as we shall discuss in more detail. While the PC2, PC3 and PC4 loadings from analyzing this dataset  are fluctuations around $\phi=\pi/2$.  With the column-centered data, all PC loadings are measuring the fluctuations. The PC1, PC2 and PC3 in this case are similar to the PC2, PC3 and PC4 from analyzing the raw data as one expects. The large values of these PC loadings in several bins around $\phi=\pi/2$ indicate some collective pattern of fluctuations of the squeezed-out nucleons. These PC loadings show some dependences on the EOS used, but there appears no clear pattern. While the standarized dataset (column-centered data scaled by the standard deviation of each bin) measures the relative event-by-event fluctuation with respect to the event averaged fluctuation from bin to bin.  Since each element in the data matrix in this case is a ratio of two fluctuations, the PC loadings from this analysis are not expected to show any collectivity but relatively large statistical fluctuations. Moreover, since nucleons at mid-rapidity with high-transverse momentum are mostly being squeezed-out perpendicular to the reaction plane, there are few particles in the reaction plane. Thus, in this case, the statistical errors of the PC loadings near the reaction plane (i.e., with $\phi$ near 0 and $\pi$) are larger. Most strikingly, none of the PC loadings in all three analyses are naturally sine and/or cosine functions. 

The corresponding event averaged PC coefficients and scaled singular values (scaled by the singular value of PC1) are shown in Fig. \ref{b6CE}, and Fig. \ref{b6SV}, respectively.  In the case of using the raw data, the singular value is overwhelmed by the contribution from PC1, while all the higher order PCs contribute much less but almost equally as shown in Fig. \ref{b6SV}. Consequently, only the coefficient of PC1 (in expressing the event averaged azimuthal angle distribution $dN/d\phi$ in terms of the PCs) is relevant in this case, and the coefficients of all higher PCs are essentially zero as shown in the left panel of Fig. \ref{b6CE}.
On the other hand, in the case of using the column-centered dataset, the singular value decreases slowly and shows a clear EOS dependence. While in the case of using the standarized dataset, the singular value does not decrease much until the last few PCs  (i.e., no ``principal components"). In both cases, the PC coefficients are all roughly equally small. Since the major expected benefit of doing PCA for some dataset is the dimension reduction. The very slow decreases of the singular values in the last two analyses presented above indicate that such benefit is not realized in the PCA-SVD analysis of fluctuations in the nuclear azimuthal flow.

\begin{figure*}[thb]
\centering
\includegraphics[width=1.\textwidth]{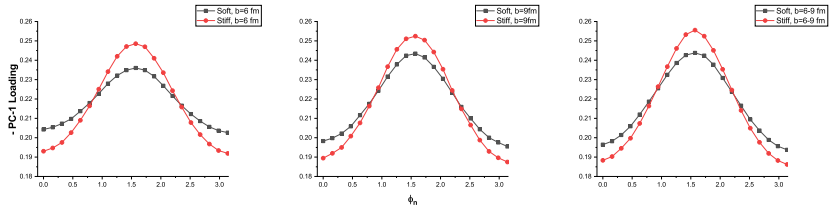}
\caption{A comparison of the PC1 loadings with an impact parameter of 6 fm (left), 9 fm (middle) and mixed evets (right) in the case of all using the raw dataset.}\label{pc1s}
\end{figure*}
 
Results of using an impact parameter of b=9 fm are qualitatively consistent with those using b=6 fm, except that the statistical errors in the case of using the standardized dataset are larger as one expects.  For the 1.5M mixed events with the impact parameter b between 6 fm and 9 fm, as shown in Fig. \ref{bmixed},  the main features are the same as for the b=6 fm case, except now with larger statistical errors in the case of using the standardized dataset. This is expected as in the reaction with b=9 fm, a lot less free nucleons are emitted, and thus the relative statistical errors are larger in the mixed events where the reactions with b=9 fm weights more than with b= 6 fm.  

As noticed earlier, in the case of using the raw dataset, the first PC loading just reflects the average azimuthal angle distribution. It would thus be interesting to examine more quantitatively its EOS dependence and how it compares with the result shown in Fig. \ref{dndf}. For this purpose, shown in Fig. \ref{pc1s} is a comparison of the PC1 loadings (multiplied by a minus sign) with an impact parameter of 6 fm (left), 9 fm (middle) and mixed events (right) in the case of all using the raw dataset. Indeed, its main features and EOS dependences are qualitatively very consistent with those shown in the average azimuthal angle distribution shown in Fig. \ref{dndf}. 

\begin{figure*}[thb]
\centering
\includegraphics[width=0.9\textwidth]{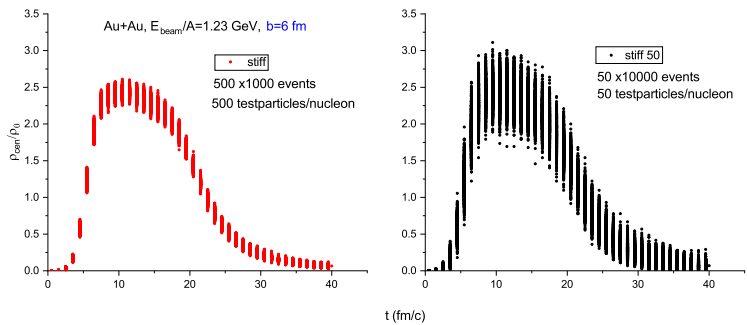}
\vspace{-5cm}
\caption{Left: Time evolution of the central density in Au+Au collisions at $E_{\rm beam}/A$=1.23 GeV using the stiff EOS with 500 testparticles/nucleon and 1,000 random impact parameters between b=6.0 fm and 6.1 fm. Right: the same as left but with 50 testparticles/nucleon  and 10,000 randow impact parameters in the same range.}\label{densityP}
\end{figure*}

\begin{figure}[thb]
\centering
\includegraphics[width=0.5\textwidth]{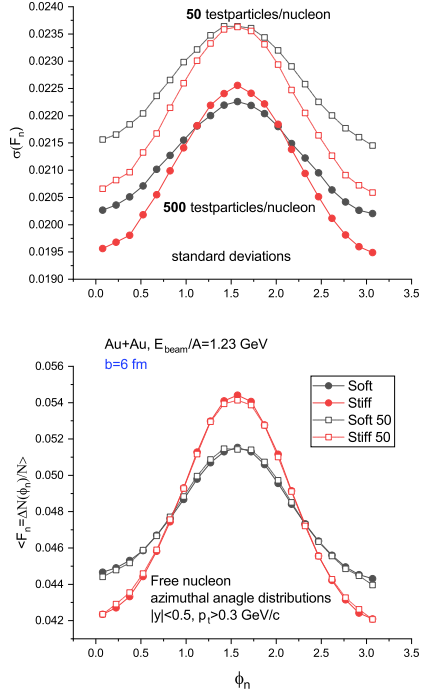}
\caption{The mean values (lower panel) and standard deviations (upper panel) of free proton azimuthal angle distributions as in Fig. \ref{dndf} (which used 500 testparticles/nucleon (soft and stiff on solid lines)) but including new results obtained using 50 testparticles/nucleon (soft 50 and stiff 50 on short-dot lines).} \label{dndf2}
\end{figure}
\section{Effects of the number of testparticles used in the IBUU transport model simulations}\label{Ntest}
All results presented above are obtained from using the default 500 testparticles/nucleon in the IBUU transport model simulations. As mentioned earlier, it means that the nucleon spatial density distribution, time evolution and the corresponding mean-field potential are averaged over 500 parallel ensembles during nucleus-nucleus collisions. While stochastic collisions among particles within different ensembles make the final states of these ensembles different, the 0.5M events of ``fixed" impact parameter of b=6 fm (1,000 different b values randomly generated between 6.0 and 6.1 fm) are not all completely independent. Namely, our ``event-by-event" flow analysis has to be taken with some cautions. As we have demonstrated above, the higher-order PC loadings above PC1 in the raw dataset and all PC loadings in the column-centered dataset measure the collectivity of azimuthal flow fluctuations. It is thus necessary and interesting to know how the number of testparticles/nucleon we us may affect our conclusions. To answer this question, we have generated two more sets of 0.5M events each with the same soft and stiff EOSs but now 50 testparticles/nucleon and 10,000 (instead of the previously used 1,000) impact parameters between 6.0 and 6.1 fm. We purposely keep the same total number (0.5M) of quasi-independent events. Shown in Fig. \ref{densityP} is a comparison of the time evolution of the central baryon density in Au+Au collisions at $E_{\rm beam}/A$=1.23 GeV in our default calculation (left panel with 500 testparticles/nucleon and 1,000 random impact parameters between 6.0 and 6.1 fm) and the new calculation (right panel with 50 testparticles/nucleon  and 10,000 random impact parameters in the same range. While a time step of 0.5 fm/c is used in the IBUU simulations, the central density at every 1 fm/c is shown to reduce the data size in the plot. Each point on any vertical line at a given instant corresponds to a specific impact parameter b between 6.0 fm and 6.1 fm. As expected, with 50 testparticles/nucleon, the central density fluctuates in a larger region compared to the default case. It is seen that the event averaged central density is about the same in both cases over the whole reaction period (50 testparticles/nucleon is approximately the minimum for this to remain true with the design of the density matrix using 1 fm$^3$ cells in the IBUU code). We notice that the maximum density formed in these reactions is about $2.3\rho_0$ around t=13 fm/c.

\begin{figure*}[thb]
\centering
\includegraphics[width=0.45\textwidth]{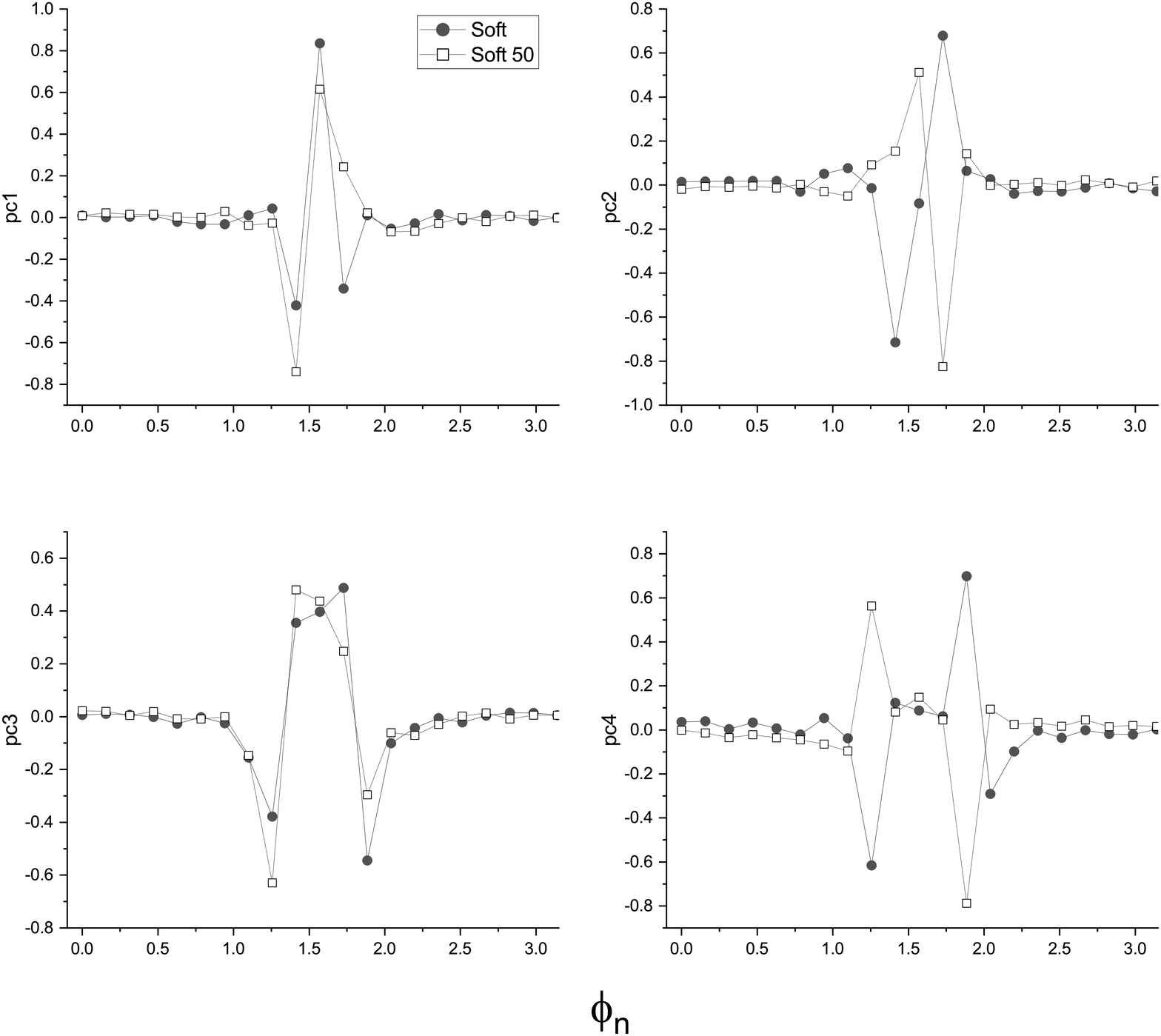}
\includegraphics[width=0.45\textwidth]{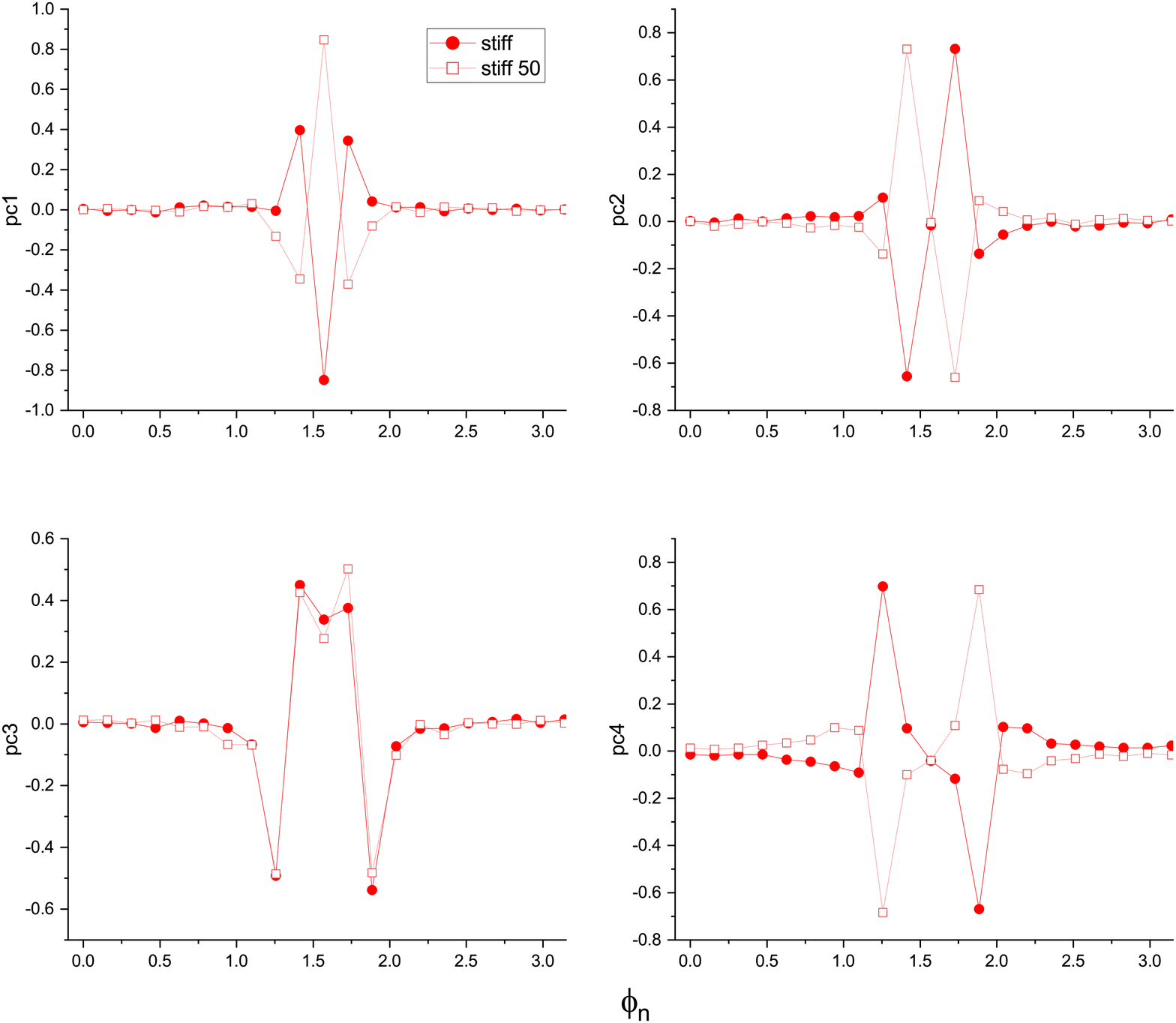}
\caption{A comparison of the first 4 PC loadings using the column-centered dataset with 500 (solid lines) and 50 (dotted lines) testparticles/nucleon using the soft (left) and stiff (right) EOS, respectively.}\label{pcsoft}
\end{figure*}
The mean values and standard deviations of the azimuthal angle distributions using 50 testparticles/nucleon are compared to the default calculations using both the soft and stiff EOSs in Fig. \ref{dndf2}. It is seen that the mean values (lower panel) of the azimuthal angle distributions remain almost the same with either 50 or 500 testparticles/nucleon as one expects. On the other hand, it is interesting to see in the upper panel that the standard deviations increase by about 5\% with the reduced number of testparticles/nucleon. Moreover, because of the smaller number of free nucleons and the weaker elliptical flow with the soft EOS, the resulting standard deviations increase more than that from using the stiff EOS especially around $\phi=\pi/2$ when the number of testparticles/nucleon is reduced to 50.  Thus, for the same number of quasi-independent events, the fluctuations of the azimuthal angle distribution do change appreciably depending on the number of testparticles/nucleon used in the IBUU transport model.

Since the density fluctuation depends on the number of testparticles/nucleon used and its dynamical effect on the azimuthal flow depends on the stiffness of the EOS, the resulting fluctuations of azimuthal flow may depend on both the EOS and the testparticles/nucleon used. To check this speculation, shown in Fig. \ref{pcsoft} is a comparison of the first 4 PC loadings using the column-centered dataset with 500 (filled symbols on solid lines) and 50 (open symbols on dotted lines) testparticles/nucleon using the soft (left) and stiff (right) EOS, respectively. Recalling that the signs of the eigenvectors are arbitrary, it is the patterns of the eigenvectors that carry useful information. The obvious phase changes in the fluctuation modes (e.g., pc2 and pc4 with the soft EOS, and pc1, pc2 and pc4 with the stiff EOS) due to the change in testparticles/nucleon can be removed by flipping the sign of one of the two eigenvectors (corresponding to exchanging $+p_x$ and $-p_x$). The resulting collective patterns of the azimuthal fluctuations around $\phi=\pi/2$ are not much affected by the testparticles/nucleon. As mentioned before, effects of the EOS on these PC loadings are weak.

\section{Summary and discussions}
Using IBUU transport model generated events for mid-central Au+Au collisions at $E_{\rm beam}/A$=1.23 GeV,
we performed SVD-PCA analyses of the azimuthal angle distributions of free nucleons for the purpose of extracting information about the EOS of dense matter formed in the reaction. We also examined whether the SVD-PCA analyses can prove that the harmonic functions constitutes naturally the most optimal basis for flow analyses using the three different ways of preparing the data matrices. We found a negative answer to the last question. Moreover, because the EOS effects on the azimuthal distribution function $dN/d\phi$ are being shared by the PC loadings, singular values and event averaged coefficients, they all show weaker dependence on the EOS while in the Fourier analyses all EOS information is carried by the harmonic coefficients. In particular, the strength of elliptical flow has the strongest sensitivity to the varying EOS. 

Our studies in this work have caveats and there are interesting issues left for further studies. Chief among them are : (1) the BUU transport model has its limitations in describing correlations/fluctuations of heavy-ion collisions, (2) it is unclear why the Harmonic functions are not found naturally to be the most probable basis of the azimuthal angle distribution in heavy-ion collisions at intermediate energies unlike what were found in Refs.  \cite{Liu19,Alt19} for LHC/RHIC energies, and (3) the relative advantages and disadvances of Fourier and PCA analyses of azimuthal flow at different beam energies are not completely clear.  While we can not fully address all of these issues at this time, based on our discussions with some experts \cite{UO} and comments by the referee of this manuscript we summarize our current understanding about these issues in the following:
\begin{itemize}

\item It is well known that BUU transport models may lack true multi-particle correlations. Nevertheless, the BUU approach has been used successfully in describing the average dynamics of heavy-ion collisions at intermediate energies, in a way similar to how hydro describes LHC/RHIC collisions even though it only propagates single particle distributions. This approach may be problematic if one wants to discuss higher order flow coefficients which are caused by initial fluctuations of the eccentricity (as at the LHC/RHIC) which would likely not be properly captured by the BUU approach. This can have consequences on our conclusions. 

\item While the PC1 from analyzing the raw dataset shows clearly strong bin-by-bin correlation signatures of elliptical flow as it measures essentially the mean value of $dN/d\phi$, the fluctuations of $dN/d\phi$ measured by higher order PC loadings appear only in a few bins around $\pi/2$.  These fluctuations do not behave harmonically and their singular values do not drop quickly with the increasing number of PCs. This is likely a consequence of the lack of correlations among fluctuations of different bins. If there were higher correlations among them then the projection PCA makes onto a lower dimensional space would yield quickly decreasing singular values which would demonstrate that fewer principal components are needed to explain a similar level of variance. Moreover, if the structure of the correlation among bins was periodic then one may see a Fourier series; however, this requirement on the inter-bin correlation of fluctuations is quite strong. 

\item Because the density matrices used in the BUU to calculate the mean-field potentials are averaged over the number of testparticles/nucleon used, only quasi-independent events can be generated in the BUU approach. While we found only appreciable changes in the PC eigenvectors and the standard deviations but not the mean values of the nucleon azimuthal angle distributions when the testparticles/nucleon is reduced from 500 to 50, our flow analysis is not fully event-by-event. In fact, it seems doubtful that a full event-by-event analysis would work for this low beam energies, especially since experiments usually have limited acceptance and efficiency. Such kind of event-by-event study is fruitful for the LHC/RHIC where there are on the order of 1000 charged particles per event. This problem has also been addressed e.g. in a recent study that employed a Deep Neural Network to classify the EOS from event-by-event data generated by using UrQMD for future CBM experiments at FAIR \cite{Om}.

\item While it seems to us that the Fourier analysis is more useful for extracting reliable information about the EOS of dense matter formed in heavy-ion collisions at intermediate energies, recent studies showed that for these beam energies, the higher order flow coefficients can be purely explained as products of the lower ones, i.e. the higher ones are fully correlated with the lowest one \cite{fr1}. Therefore, one may argue that the PCA finds the Fourier series at LHC/RHIC energies because the different flow coefficients are essentially uncorrelated and related to initial state fluctuations. At lower energies, however, the different orders are strongly correlated and thus the Fourier series is not orthogonal and is therefore not identified by the PCA. 

\item While we have found indications that the traditional method of flow analysis works well for extracting information about the EOS, there may be an argument on the contrary. If all flow coefficients are more or less correlated at low beam energies, then no new information can be gained by measuring more than one of them, even though the sensitivity is still large. The PCA supposedly gives independent/uncorrelated observables. So having 3 flow coefficients which are correlated vs. 3 PCs which are uncorrelated is different and one can make the argument that the latter has more benefits. Hopefully, the results presented here provide a useful basis for further exploring the relative advantages and distadvances of Fourier and PC analyses of azimuthal flow in heavy-ion collisions from low to ultra-relativistic beam energies. 

\end{itemize}

\section*{Acknowledgement}
This work is supported in part by the U.S. Department of Energy, Office of Science,
under Award No. DE-SC0013702, the CUSTIPEN (China-
U.S. Theory Institute for Physics with Exotic Nuclei) under
US Department of Energy Grant No. DE-SC0009971.
We would like to thank  U. W. Heinz and J.-Y. Ollitrault for very useful discussions on the SVD-PCA of azimuthal flow in heavy-ion collisions at different beam energies during the INT program on the Intersection of Nuclear Structure and High Energy Nuclear Collisions, and Xian-Gai Deng, Yu-Gang Ma, Wen-Jie Xie and Kai Zhou for very helpful discussions on machine learning techniques. 
\\\\
 \noindent{{\bf Appendix}: Integrated and differential transverse and elliptical flow of free nucleons in mid-central Au+Au collisions at $E_{\rm beam}/A$=1.23 GeV}\label{old-flow}

The transverse flow (also called directed flow) 
has been studied most commonly by analyzing the average 
transverse momentum per nucleon in the
reaction plane as a function of rapidity $y$ \cite{pawel85}
\begin{equation}
< p_x/A >(y)=\frac{1}{A(y)}\sum_{i=1}^{A(y)} p_{ix}(y)
\end{equation}                                                                                                                                                                                                      
where $A(y)$ is the number of nucleons at rapidity $y$ and $p_{ix}(y)$ ($i=1$ to $A(y)$) are their transverse momenta in the $x-$direction. It is also frequently referred as integrated (over $p_t$) transverse flow. In setting up the simulations, we use the convention that $<p_x>$ (and the corresponding $v_1$) is positive for forward (positive $y$) going particles in the center of mass (cms) frame of the two colliding nuclei.
\begin{figure}[thb]
\centering
\includegraphics[width=0.5\textwidth]{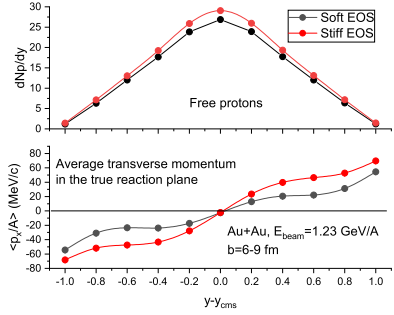}
\caption{The rapidity distribution (upper) and average in-plane transverse momentum $<p_x/A>$ (lower) of free protons in mid-central Au+Au collisions at $E_{\rm beam}/A$=1.23 GeV with a soft (K=230 MeV) and a stiff (K=380 MeV) EOS, respectively.} \label{tflow}
\end{figure}

Shown in Fig. \ref{tflow} are the rapidity distribution (upper) and average in-plane transverse momentum $<p_x/A>$ (lower) of free protons in the mid-central Au+Au collisions at $E_{\rm beam}/A$=1.23 GeV with the soft and stiff EOS, respectively. The main features including the EOS dependence of these distributions are all well understood and are consistent with previous studies. They are presented here for completeness and comparisons with the PCA-SVD analysis for the purpose of extracting the EOS information. Normally one uses the slope of the $<p_x/A>$ at mid-rapidity and/or its magnitude around the target/projectile rapidity to characterize the strength of directed flow. It is seen that they both show significant EOS effects. 

The differential directed flow as a function of $y$ and $p_t$ is characterized by the strength of the first harmonics 
\begin{equation}
v_1(y,p_t)=<cos(\phi)>(y,p_t)=\frac{1}{n}\sum_{i=1}^{n}\frac{p_{ix}}{p_{it}}
\end{equation}
while the differential elliptical flow is described by 
\begin{equation}
v_2(y,p_{t})=<cos(2\phi)>(y,p_t)=\frac{1}{n}\sum_{i=1}^{n}\frac{p_{ix}^2-p_{iy}^2}{p_{it}^2}
\end{equation}
where $n(y,p_t)$ is the total number of particles with rapidity $y$ and transverse momentum $p_t$. 
Compared to the integrated flow, the differential flows may help uncover more detailed information about the EOS of dense matter, see, e.g., Ref. \cite{Li-sustich}. On the other hand, the $v_1(y,p_t)$ and $v_2(y,p_t)$ normally depend strongly on the rapidity $y$ and transverse momentum $p_t$, requiring detailed analyses and the results normally have strong dependences on the acceptances of the detectors. 

\begin{figure}[thb]
\centering
\includegraphics[width=0.55\textwidth]{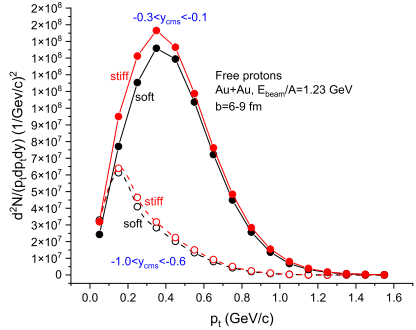}
\caption{The transverse momentum distributions 
$d^2N/(p_tdp_tdy)$ of free protons in the two representative rapidity windows indicated for the mid-central Au+Au reactions.} \label{ptd}
\end{figure}

Before analyzing the $v_1(y,p_t)$ and $v_2(y,p_t)$, it is instructive to first examine the transverse momentum distributions 
$d^2N/(p_tdp_tdy)$ in two representative rapidity windows. Shown in Fig. \ref{ptd} are the $d^2N/(p_tdp_tdy)$ of free protons in the Au+Au reactions in the rapidity range of $0.1\leq |y_{\rm cms}|\leq 0.3$ and $0.6\leq |y_{\rm cms}|\leq 1.0$, respectively. Because of the symmetry of the reaction, as shown in Fig.\ref{tflow},
these two rapidity windows contain typical nucleons from the participants and target/projectile spectators. It is seen that 
around the target/projectile rapidity $(\pm 0.74)$ in the range of $0.6\leq |y_{\rm cms}|\leq 1.0$ where the $<p_x/A>$ is the strongest, the $d^2N/(p_tdp_tdy)$ peaks around $p_t=0.15$ GeV/c. While in the participant region in the rapidity range of $0.1\leq |y_{\rm cms}|\leq 0.3$, the $d^2N/(p_tdp_tdy)$ peaks at a much higher value around $p_t=0.15$ GeV/c. Moreover, in this rapidity range, the $d^2N/(p_tdp_tdy)$ shows a significantly stronger EOS effect. Interestingly, the EOS effects are actually most visible around and/or below the peaks of $d^2N/(p_tdp_tdy)$ instead of in its high-momentum tails. This information might be useful for designing experiments searching for EOS effects. Overall, the EOS information revealed from analyzing the $d^2N/(p_tdp_tdy)$ versus $p_t$ and the $<p_x/A>$ versus $y$ are consistent and complementary to each other.

\begin{figure}[thb]
\centering
\includegraphics[width=0.55\textwidth]{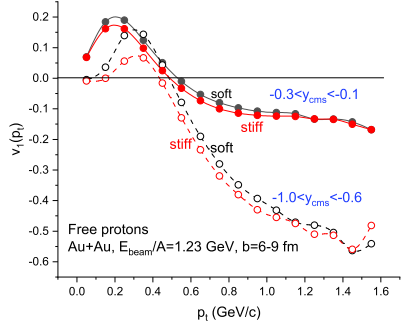}
\caption{ The differential directed flow $v_1(y,p_t)$ values as functions of $p_t$ in the specified rapidity ranges for the mid-central Au+Au reactions.}\label{v1pty}
\end{figure}
We now turn to the differential transverse flow $v_1(y,p_t)$. Shown in Fig. \ref{v1pty} are the $v_1(y,p_t)$ values as functions of $p_t$ for the near mid-rapidity and target/projectile rapidity ranges. First, it is seen that there is a change in sign around $p_t\approx 0.5$ GeV/c. The majority of free protons in the low $p_t$ region (thus also low energy $E=\sqrt{m^2+p_t^2}{\rm cosh(y)}$) have positive $p_x$ (thus also $v_1$), while the high energy protons have negative $p_x$. The net sums of these particles lead to the negative $<p_x/A>$ in the two rapidity ranges considered. 
Considering the information from the $p_t$ distribution in Fig. \ref{ptd} and the average in-plane transverse momentum in Fig.\ref{tflow}, it is seen that it is the high-momentum nucleons dominate the $v_1(y,p_t)$. In particular, around the target/projectile rapidity, this phenomenon is stronger. While there are only few high-$p_t$ free protons as shown in Fig. \ref{ptd}, the large $p_x$ values carried by these high-$p_t$ particles contribute more to the net $<p_x/A>$ compared to the contributions of a lot randomly moving low-$p_t$ particles. As to the effects of nuclear EOS, the integrated directed flow $<p_x/A>$ appears to be a better tool compared to the differential one $v_1(y,p_t)$ although they bare consistent EOS information.

\begin{figure}[thb]
\centering
\includegraphics[width=0.55\textwidth]{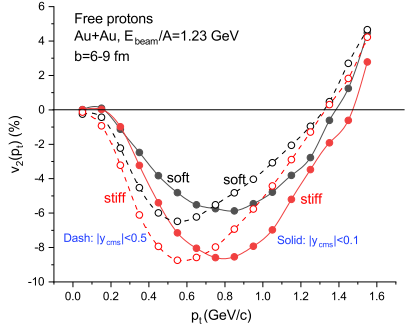}
\caption{The differential elliptical flow $v_2(y,p_t)$ values as functions of $p_t$ in the specified rapidity ranges for the mid-central Au+Au reactions.}\label{v2pty}
\end{figure}
Shown in Fig. \ref{v2pty} are the differential elliptical flow $v_2(y,p_t)$ values as functions of $p_t$ in the two mid-rapidity bins for the mid-central Au+Au reactions. While the low-$p_t$ free protons are azimuthally isotropic, the ellipticity increases to about -9\% (the squeeze-out perpendicular to the reaction plane dominates over the in-plane flow) at $p_t\approx 0.6\sim 0.8$ GeV/c. It then decreases and finally change sign at very high-$p_t$ (in-plane flow dominates). Compared to both the integrated and differential transverse flows studied above, it is interesting to see clearly that the elliptical flow $v_2(y,p_t)$ has the strongest sensitivity to the variation of nuclear EOS. Moreover, the sensitivity increases slightly when the rapidity range is further narrowed towards the mid-rapidity. Of course, because of the total energy-momentum conservation, the ellipticity peaks at different $p_t$ values in the two rapidity ranges considered. 



\end{document}